\documentstyle[prl,aps]{revtex}

\begin{document}
\baselineskip=12pt

\draft
\flushbottom
\twocolumn[\hsize\textwidth\columnwidth\hsize
\csname@twocolumnfalse\endcsname
\title{Magnetoresistance in Mn pyrochlore: electrical transport
in a low carrier density ferromagnet}
 
\author{Pinaki Majumdar and Peter Littlewood} 
\address{ Bell Laboratories, Lucent Technologies, 600 Mountain Ave,
 Murray Hill, NJ 07974}

\date{\today}

\maketitle
\tightenlines
\widetext
\advance\leftskip by 57pt
\advance\rightskip by 57pt
\begin{abstract}

We discuss magnetotransport in a low density electron gas coupled 
to spin fluctuations near and above a ferromagnetic transition. 
Provided the density is low enough ($n \lesssim 1/\xi^{3}(T)$,
with $\xi(T)$ the ferromagnetic correlation length), spin 
polarons form in an intermediate temperature regime above $T_c$.
Both in the spin polaron regime, and in the itinerant regime 
nearer $T_c$, the magnetoresistance is large. We propose that
this provides a good model for ``colossal'' magnetoresistance 
in the pyrochlore Tl$_{2-x}$Sc$_x$Mn$_2$O$_7$, fundamentally 
different from the mechanism in the perovskite manganites such 
as La$_{1-x}$Sr$_x$MnO$_3$. 

\

\

\end{abstract}

]

\narrowtext
\tightenlines

In recent years ``colossal magnetoresistance'' (CMR), particularly 
in the perovskite manganite La$_{1-x}$Sr$_x$MnO$_3$ and its
 variants, has emerged as a rich and extremely active area of
 experimental study~\cite{cmrexpts,cmrref}.
The phenomena of magnetic transition and the simultaneous 
insulator-metal transition, as the temperature is lowered, is
qualitatively understood as arising out of a combination of
Mn$^{3+}$-Mn$^{4+}$ double-exchange and transport via Jahn-Teller 
polarons~\cite{millis,roder}.
The magnetic exchange arises from electron hopping, itself 
dependent 
on the spin order, while Jahn-Teller distortions and the atomic
size mismatch between Mn$^{3+}$ and Mn$^{4+}$ trap electrons in
small polaronic states. The magnetic transition involves 
the cooperative effect of both the charge and spin degrees of
freedom; spin ordering promotes electron hopping, increases the
effective exchange, anneals out the lattice distortions and, in a 
bootstrap effect, leads to the magnetic and insulator-metal
transition.  

The pyrochlore Tl$_2$Mn$_2$O$_7$ offers a surprising contrast, and
demonstrates that neither double exchange nor lattice polarons are  
{\it essential} for obtaining CMR.
From recent experiments~\cite{shimikawa,cheong,sc1,sc2} the
following picture has emerged.
As in the perovskites, the large MR accompanies a paramagnet to 
ferromagnet transition, with $T_c$ around 140K. However, the carrier
density estimated from Hall effect is low~\cite{shimikawa}
($\sim$ 0.001 - 0.005 per formula unit, f.u.), and the
ferromagnetic transition is driven by superexchange between the
Mn sites, close to their nominal valence of Mn$^{4+}$~\cite{sc1}.
The thermopower~\cite{sc2} in Tl$_2$Mn$_2$O$_7$ is almost two
orders of magnitude larger than in good metals, attesting to a 
low Fermi energy.
Electrical transport is provided by carriers on the Tl sites, in 
the tail of a broad Tl-O hybridised band that may overlap the
topmost Mn d band~\cite{bandstruct}.
{\it There is thus neither double exchange inducing the magnetism, 
nor a driving mechanism for lattice polarons or Jahn-Teller
distortions.}

Nevertheless the MR is very large above $T_c$ even though the 
resistance is
``metallic'' ($d\rho / dT > 0$) in the paramagnetic phase for 
$T \gtrsim 1.5 T_c$.  In terms of the
rough scaling relation $(\rho (0)-\rho (H))/ \rho (0)
\approx C (m/m_s)^2$
above $T_c$ ($m$ and $m_s$ are the magnetization and saturation 
magnetization respectively), the coefficient $C \approx 15$ is
even larger
than observed~\cite{urushibara} in the  metallic  perovskite 
manganites.
With substitution by In~\cite{cheong}
or Sc~\cite{sc2} on the Tl site, the magnetic properties
are weakly affected, while the transport is dramatically modified. 
The resistivity increases by orders of magnitude~\cite{sc2}, 
becomes activated in the paramagnetic phase and the MR 
increases further. 

This paper  argues that the data provoke a simple model of a low 
density electron gas interacting with a spin background that orders 
ferromagnetically, {\em independently} from the conduction
electrons.
Although the density is low enough that the average
magnetic properties (e.g. $T_c$) are hardly affected by the
carriers, 
at low enough density, and sufficiently large electron-core spin 
coupling, carriers will self trap into well defined, non
-overlapping, magnetic polarons. 
The core size of the magnetic polaron
increases with decreasing temperature  remaining finite at $T_c$, 
but the ``interface'' width, over which the local magnetisation 
decays, is the magnetic correlation length, $\xi (T)$,  which
diverges as $ T \rightarrow T_c$.
When the density $n \approx \xi^{-3}$ (see Fig.1) 
the polarons overlap and the carriers delocalise.
In both the itinerant and the self-trapped regime we find the MR
to be large.

To be specific, we consider the Hamiltonian
\begin{eqnarray}
\hat H&=&  \sum_ {{\vec k},{\sigma}} (\epsilon_{\vec k} - \mu)
c^{\dagger}_{{\vec k},{\sigma}}
c_{{\vec k},{\sigma}}  
-J'\sum_i
{\vec \sigma}_i \cdot {\vec S}_{i}  \nonumber \\
% \cdot \sum_{\vec R = n.n.}{\vec S}_{i+ {\vec R}}  \nonumber \\
&-& J\sum_{\langle i,j \rangle} {\vec S}_i \cdot {\vec S}_j
- \sum_i {\vec h} \cdot {\vec S}_i
\end{eqnarray}
Here $S_i$ refer to the localised Mn core spin (S=3/2), 
and $J$ sets the scale for $T_c$ (mean field $T_c \sim zJS^2$,
and $z$ is
the Mn coordination).
$c$,$c^{\dagger}$ refer to carriers in the Tl-O band\cite{shift}
and 
${\vec \sigma}_i=
c^{\dagger}_{i,\alpha}
{\vec \sigma}_{\alpha,\beta}
c_{i,\beta} $ is the conduction electron spin operator.
$J'$ is the effective exchange coupling between a Mn spin and the
conduction electron, and $h$ is the external field. 

For the Mn pyrochlores, we expect that 
$t \sim {\cal O}(0.1)$ eV~\cite{bandstruct}, and $J' / t $ may be
of order unity~\cite{exchange}.
The transition temperature $T_c \sim 140K$.
The carrier density in the nominally
undoped compound is $\sim 10^{-2}-10^{-3}$ /f.u, while in the
Sc doped
systems~\cite{sc2}  the combined effect of disorder and lowered 
carrier density can be
inferred from $\rho(T)$ as  $T\rightarrow 0$.

%
%	fig 1
%

Since our principal goal is to understand  transport properties,
and our assumption is that the spin correlations are 
{\em on average} unaffected by the carriers, we shall take the
spin correlations to be given by the ferromagnetic Heisenberg model.
In practice we shall use mean field theory and  Ginzburg-Landau (GL) or 
Ornstein-Zernicke (OZ)
approximations  for the correlation functions, since
we are not concerned with details in the vicinity of $T_c$.
We need to consider transport in the two regimes of Fig.1, and
begin with the itinerant regime.

Fluctuations near any critical point usually lead to large
scattering but  the dominant $q \rightarrow 0$
fluctuations near a ferromagnetic transition usually have a 
negligible
effect on transport because it is primarily modes near $q\sim 2k_F$
which are effective in backscattering. The obvious and interesting
exception is a low electron density system, $k_Fa \ll 1$ ($a$ is the
lattice constant), where
the growth of magnetic fluctuations can be  directly reflected
in the resistivity. 
The standard theory for the ``spin disorder'' contribution to
resistivity near a ferromagnetic transition 
was  given  by de Gennes and Friedel~\cite{dgfr}, subsequently
criticised and modified by Fisher and Langer~\cite{fl}. 
This Born scattering result for the transport relaxation rate
$\tau^{-1}$, normalised to its high temperature value $\tau_0^{-1}
\sim (J'^2/t)S(S+1)k_Fa $,
is given by
$$
\tau^{-1}/\tau_0^{-1}
\sim \int_0^{\pi} \sigma(\theta) (1-cos \theta)sin \theta d\theta
$$
where $\sigma(\theta)$ is the differential scattering cross section per
magnetic spin, and $\theta$ the scattering angle. 
The cross section, in turn, is given by
$$
\sigma(\theta) \equiv \sigma(q=2k_F sin (\theta/2)) \sim \chi(q)
$$
where $\chi(q)$ is the static structure factor. Within the 
OZ
approximation $\chi(q) \sim \xi^2/(1 + q^2 \xi^2)$,
where $\xi$ is the magnetic correlation length.
The form for
$\tau^{-1}$ is easy to evaluate using $\chi(q)$ above but a fair
amount of insight can be gained by
simply using $\tau^{-1} \sim \chi(q\sim 2k_F)$.
This is featureless for $k_Fa \sim  {\cal O}(1)$, but picks up
 significant
temperature dependence for $k_Fa \ll 1$, with $\tau^{-1}/\tau_0^{-1}
\sim (k_F^2a^2 + T/(T-T_c))^{-1}$.

The complete answer for the scattering rate, within the OZ 
approximation, 
is
\begin{equation}
\tau^{-1}/\tau_0^{-1}
 \sim {1 \over {k_F^2a^2}} (4 - {1 \over {k_F^2 \xi^2}} \log
(1 + 4k_F^2 \xi^2))
\end{equation}
This result should be modified close to $T_c$,
where non mean-field effects are important, and also
when $\xi(T) \gtrsim l(T)$, the mean free path~\cite{fl}. 
These effects remove the cusp-like $T$-dependence at $T_c$, 
but the important density-dependence remains unchanged.
Notice that since $\chi(0) \sim 
\xi^2$ within the OZ theory, Eq.~2 implies a direct relation 
between the scattering rate and the 
susceptibility. For $k_F \xi(T) \ll 1$ it is easy to see that
$\tau^{-1} \sim \xi^2 \sim \chi$ over a wide temperature range
emphasising that $d\rho/dT <0$ is possible in the paramagnetic
{\it metallic} state.

We calculate the  magnetoresistance arising  from the
field suppression of magnetic fluctuations, {\it i.e}
the reduction in correlation length;
$\xi^2 \Rightarrow \xi^2(m,T)$
which, within the GL theory, can be shown to be
$\sim \partial m/\partial h\vert_m$,  
where $m(h,T)$ is the magnetisation
due to an applied field.  
We may calculate the MR from
the equation above but to make the qualitative point,
$\delta \rho/\rho(0) \sim (\tau^{-1}(m,T)-\tau^{-1}(0,T))
/\tau^{-1}(0,T)$  
which is approximately $(\chi(2k_F,m,T) - \chi(2k_F,0,T))/
\chi(2k_F,0,T)$.
Using the finite field version of $\chi(q)$ from GL, 
one can easily show that $C \sim 1/k_F^2a^2$
for $k_F\xi \gg 1$~\cite{crtscatt}. 
$C$ involves a numerical constant $\sim 1$, and 
temperature dependence arising out of $\xi(T)$, but we only 
want to
emphasise the density dependence. Obviously 
lower  densities can greatly enhance $C$
consistent with the observations in~\cite{shimikawa}, 
{\it without 
involving an insulator-metal transition}.
This perturbative framework, however, 
cannot be continued to arbitrarily low
density or to  $J'/t \gtrsim 1$
where, if the spin  background were treated as ``quenched 
disorder'',
one would expect electron localisation\cite{loc}. However, 
for  $J'/t \gtrsim 1$, and low 
carrier density,  the electrons 
actually self trap into magnetic polarons, as we discuss next.

The  issue of magnetic polarons was raised long ago~\cite{polref},
but apart from certain limiting cases studied by
 Kasuya {\it et al.}~\cite{kasuya1}
we know of no systematic calculation on the size and energy of the
bound state. 
Our  calculation consists of: (i) A  variational ansatz for the 
electron
wavefunction $\psi(r)$ (with spin $\uparrow$, say); (ii)
Calculation of the polarisation and free
energy of the spin background due to the effective ``field''
$J'\langle \sigma_z(r) \rangle$;  and (iii)
Minimisation of the total free energy; electron kinetic energy + 
magnetic free energy, with respect to the variational parameter. 
While our numerical results are shown for $S=1/2$, for simplicity,
we provide an analysis which generalises the answers to arbitrary 
$S$. 

The simplest
ansatz is that of an electron isotropically delocalised over a
region of
radius $L_p$ (measured in terms of $a$), involving $\sim L_p^3$
sites.
This leads to a ``field'' $h_p \sim J'/L_p^3$ acting on the 
spins, which lead to polarisation and  gain in magnetic 
free energy.
The magnetisation of the polarised region can be estimated 
from mean field theory, $m = tanh \beta(T_c m + h_p)$
and the mean field 
magnetic free energy is
$$
 \Delta F_m
\sim L_p^3\{{1 \over{2}} T_cm^2 - T ln(cosh\beta (T_cm + h_p))\}
$$
The total free energy $\Delta F =\Delta F_m + t/L_p^2$ is minimised
w.r.t $L_p$. Temperature dependence enters
through the magnetisation equation, which encodes the diverging
susceptibility, while external fields add to the polaronic field
and require a straightforward generalisation. Our result for
the binding energy, $\Delta_p = min ( \Delta F (L_p))$,  
as a function of temperature and 
external field is shown in Fig.2.  

Postponing a complete discussion of the polaron calculation
to a separate communication~\cite{poltext}
we remark on the essential results here.
(a).~As in~\cite{kasuya1} we find that
for a given set of parameters $\{ t, J', T_c\}$, the spin 
polaron becomes favored only {\it below a certain temperature},
$T_p$ say. Assuming a saturated core this is  approximately 
given by
$T_pln(2S+1)/t \sim (zJS^2/t) + (J'S/t)^{5/2}$. Thus the ``window'' above
$T_c$ where the polaron exists increases with $J'/t$. Fig.1
 indicates the
variation in $T_p$ with $J'/t$, deduced from the numerics, roughly
consistent with the above result.
At high temperature, the polaron is  confined
to a few sites, and
the local magnetisation is 
saturated. In fact, for $J'/t \gtrsim 1$,
$m \gtrsim 0.9$ down to $T_c$.
(b).~With reducing temperature both the polaron size 
$\bar L_p$, and $\Delta_p$ increase.
Since the numerical minimisation reveals that $m \simeq 1$
a simple analysis is possible. Close to saturation 
the magnetisation equation yields $m \sim 1 - 2e^{-2\beta (T_c
 + h_p)}$.
Using this, to leading order, 
the free energy function 
 $\Delta F \sim L_p^3(T ln2 - T_c/2) -J' +t/L_p^2$
where the terms can be readily interpreted as the magnetic free
energy of ${\cal O} (L_p^3)$  saturated spins, the ${\cal O}(1)$
 exchange energy $J'$ of the electron, and the kinetic energy. 
Minimising this  yields $\bar L_p^5 \sim (2t/3)/(Tln2 -T_c/2)$. 
The `formation' temperature is given by  $\Delta F(\bar L_p(T_p))=
0$  and 
the binding energy $\Delta_p/t \sim 
{5 \over {3}} ({ 3 \over {2t}}(Tln 2 -T_c/2))^{2/5} -J'/t$. 
This almost completely describes the numerically  obtained
zero field curve in Fig.2. (c).~In the presence of an external field
the binding energy is the {\it difference} between the energy of
 the polaron and that of the delocalised electron in the
applied field. This is  principally 
 $ \sim J'
( m - m_{ext}) $, where $m$ and $m_{ext}$ are 
respectively the core magnetisation 
and the  external magnetisation,
which diminishes as the field
magnetises the spin background. 
For fields large enough to ``saturate'' the spin  background
the magnetic energy of the carrier is $-J'$ 
irrespective of whether it is in a localised or extended state,
and the energy gain 
$\Delta_p \rightarrow 0$. Conversely, for $T \rightarrow T_c$,
when the susceptibility is largest, the reduction in binding
energy is most pronounced (Fig.2).
 
In the regime, $J'/t \sim {\cal O}(1)$, that we are 
interested in, the above analysis readily generalises to
arbitrary $S$  and we have $\Delta_p/t 
\sim {5 \over {3}}
({3 \over {2t}}(Tln(2S+1) -T_c/2))^{2/5} -J'S/t$. 
So, for a system with $S>1/2$, {\it e.g.} the pyrochlores, 
the result in Fig.2 only needs to be scaled by appropriate
factors of $S$. 

There is no accepted single theory of transport via spin polarons.
 A  ``small'' spin polaron, like the one we have computed,  with 
a large local field, presumably moves via hopping like a small
lattice 
polaron since the ``field'' would inhibit diffusion via spin flips. 
In that case  the principal mode of conduction would be polaron 
``hopping''  over a barrier or ``ionisation'' of the trapped 
carrier.
Since both these processes are activated, with energies $\sim
\Delta_p$,
one expects $ln \rho \sim \Delta_p/T$ (see Fig.2).
The large MR follows from the magnetic field dependence of 
$\Delta_p$.
Using our results for $\Delta_p(T,h)$ we estimate the  MR 
that can arise 
from an activated  transport process in Fig.3.

Before moving on to a comparison with experiments, we discuss
the regime of validity of the results and the 
approximations in the calculation.
(a).~The boundary between the polarised and
unpolarised regions is not sharp, in fact scaling as $\xi(T)$
which diverges as $T \rightarrow T_c$. 
A description in terms of isolated polarons will break down
when $n \xi^3 \approx 1$, which for the parameters used
here is in the range $T/T_c \lesssim 1.05-1.1$. 
In that regime transport would be described by
itinerant scattering, also leading to large MR (see fig. 3). 
(b).~The calculation of the bound state wavefunction 
and the magnetic polarisation should be self-consistent, and
a sharp boundary leads to an overestimate of the binding energy
for $T\rightarrow T_c, \xi \gg \bar L_p$. 
This regime, where  the electron ``delocalises'' over
a lengthscale $\sim {\cal O} (\xi)$, 
is important for $T/T_c \lesssim 1.05$.

A quantitative comparison of our results with the data on 
Tl$_{2-x}$Sc$_x$Mn$_2$O$_7$ is difficult 
because the carrier concentration is
not accurately known and disorder is not controlled. There is
substantial variation between the results of different groups on 
nominally the same material, even as to the sign of $d \rho / dT$.
However, the end member in this series, 
Sc$_2$Mn$_2$O$_7$, is a ferromagnetic insulator 
so although it is not clear how much  
the trends with Sc 
substitution~\cite{sc2} are to be ascribed to reducing carrier 
density and how much
to increased disorder, there is definitely a reduction 
in carrier density with increasing $x$.
(a).~The $T\rightarrow 0$ phase at $x=0$ is metallic, albeit with 
rather large resistivity, while for $x \gtrsim 0.2$ $\rho(T)$ as 
$T\rightarrow 0$ shows an upturn, indicating the onset
of Anderson localisation, arising out of a combination of
decreasing $n$ and increasing disorder.The resistivity
at this low-temperature metal insulator transition is compatible
with the usual Ioffe-Regel criterion. 
(b).~The resistivity in the paramagnetic phase at $x=0$ shows 
$d\rho/dT $ weakly negative  for $T \gtrsim T_c$. This, and the 
large
MR coefficient~\cite{shimikawa}, is consistent with magnetic 
scattering
in a low density metal.
( For $T > 1.5 T_c$,  $d\rho/dT >0$ probably due to non magnetic 
sources
of scattering.) 
(c).~For $x \gtrsim 0.2$ the resistivity for $T>T_c$ is
much too large to be described as a strongly scattered metal
(one would have $k_F l \ll 1$). Furthermore
$d\rho/dT <0$ upto $T \sim 350$K, and if fitted to an activated
form the activation energy is on the order of $0.1 eV$. 
 Despite the
very different absolute scales for the resistivity
 at x=0.2, 0.3 and 0.4, 
the temperature dependence is almost
identical as a normalized plot based on the data in~\cite{sc2} 
reveals.
This is a regime we believe should be described by polaronic 
transport. To compare with the data we reproduce the measured
$\rho(T)$ at $x=0.3$ in the inset of Fig.2. The
measured  magnetoresistance
for $x=0.2, 0.3$ and $0.4$ are almost identical
and we reproduce this as an inset in Fig.3 to compare with
the `MR' derived within our polaron calculation.  
The correspondence between experimental and theoretical field
scales is approximately $1$ T $\equiv 0.01 T_c$.

In conclusion, we have argued that a simple model 
scattering of carriers by ordering moments - can yield large 
MR when
the carrier density is low. At ultra-low densities, the carriers
will self trap as magnetic polarons 
and
Tl$_2$Mn$_2$O$_7$ appears to be close to this regime, especially
upon Sc substitution.
Direct evidence of spin polarons could be best sought with NMR
 and ESR
measurements, as well as the appearance of an ionization gap in the
optical conductivity.

\acknowledgements
We acknowledge discussions with Elihu Abrahams, Gabriel Aeppli,
 Bertram Batlogg, Harold Hwang, Y. B. Kim and Andy Millis.  
In particular we thank Art Ramirez for several discussions, 
a critical reading of the manuscript, and for
providing the data from~\cite{sc2} replotted in Figs.2 and 3.

FIG.1:
$n-T$ range of magnetic polarons for various $J'/t$ at $t/T_c=10$.
The window over which the polaronic description is valid 
lies between 
$\xi(T)\approx n^{-1/3}$ (when polarons overlap), and $T_p$, where the 
bound state
forms (see text). 

FIG.2: 
Binding energy $\Delta_p/T$ for 
$t/T_c=10$, $J'/t=1$  and varying  $h/T_c$.
Inset; $log \rho$ for Sc doped sample, $x=0.3$, replotted 
from~\cite{sc2} 

FIG.3:
MR for $h/T_c=0.02$, from the two scenarios.
The Born scattering result (Eqn. 2)  corresponds to $n \sim 
(k_Fa)^3 
\sim 10^{-3}$.
Inset; ``universal''
MR data in
Tl$_{2-x}$Sc$_x$Mn$_2$O$_7$ at $6$T 
for $x=0.2, 0.3, 0.4$~(replotted from~\cite{sc2}) 


\begin{references}
\bibitem{shift}This single band model does not allow for 
magnetisation dependent band shifts, relevant in a semimetal, which
maybe the case for   
 Tl$_2$Mn$_2$O$_7$. Such an effect can be straightforwardly 
incorporated as a magnetisation dependent carrier density although 
this seems not to be indicated by the data.
\bibitem{exchange}
We estimate that 
$J' \sim t_{sd}^2 ({1 \over {\Delta_{cf} -\epsilon_s -J_H}}
- {1 \over {\Delta_{cf} -\epsilon_s + J_H}})$
where $t_{sd}$ is the effective hopping between the Mn $e_g$ level
and the Tl 6s, $\Delta_{cf} -\epsilon_s$ is their energy separation, 
and $J_H$ the Hund's rule coupling of the $e_g$ level to the core
$t_{2g}$ spin $S$. $J'$  is a coupling between Tl and Mn sites,
not ``onsite'' as we have written, but that does not modify
our conclusions.
\end{references}
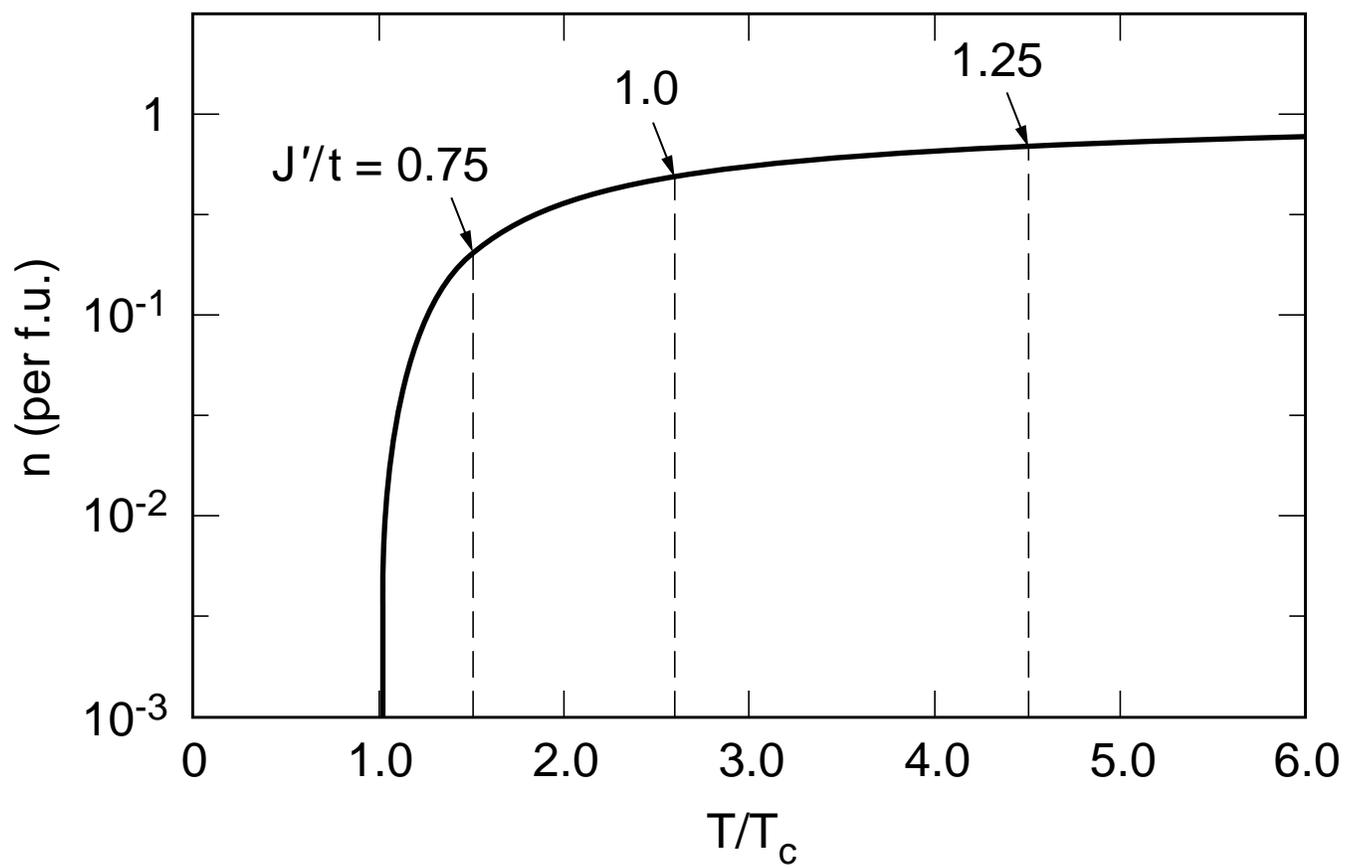
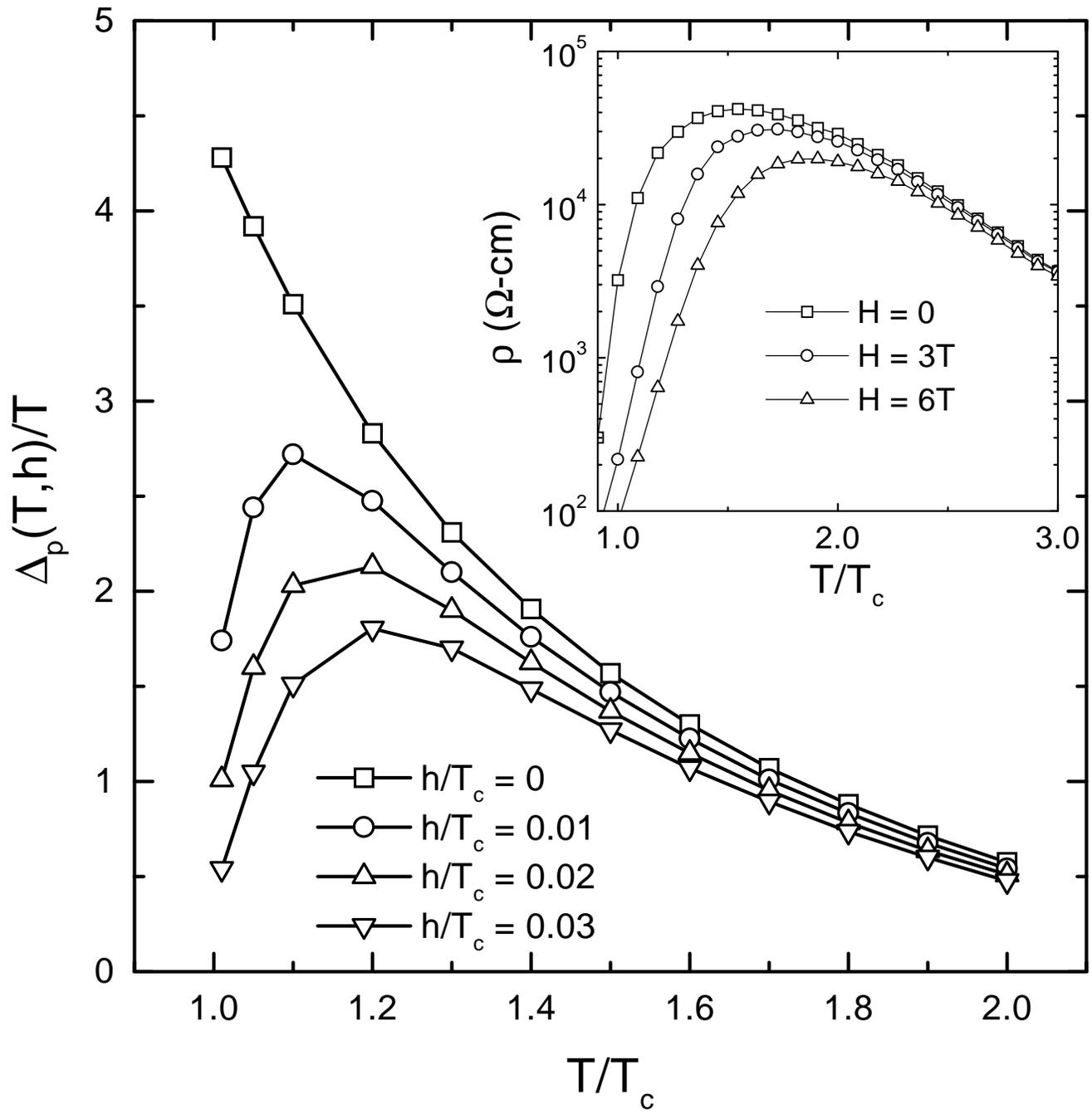
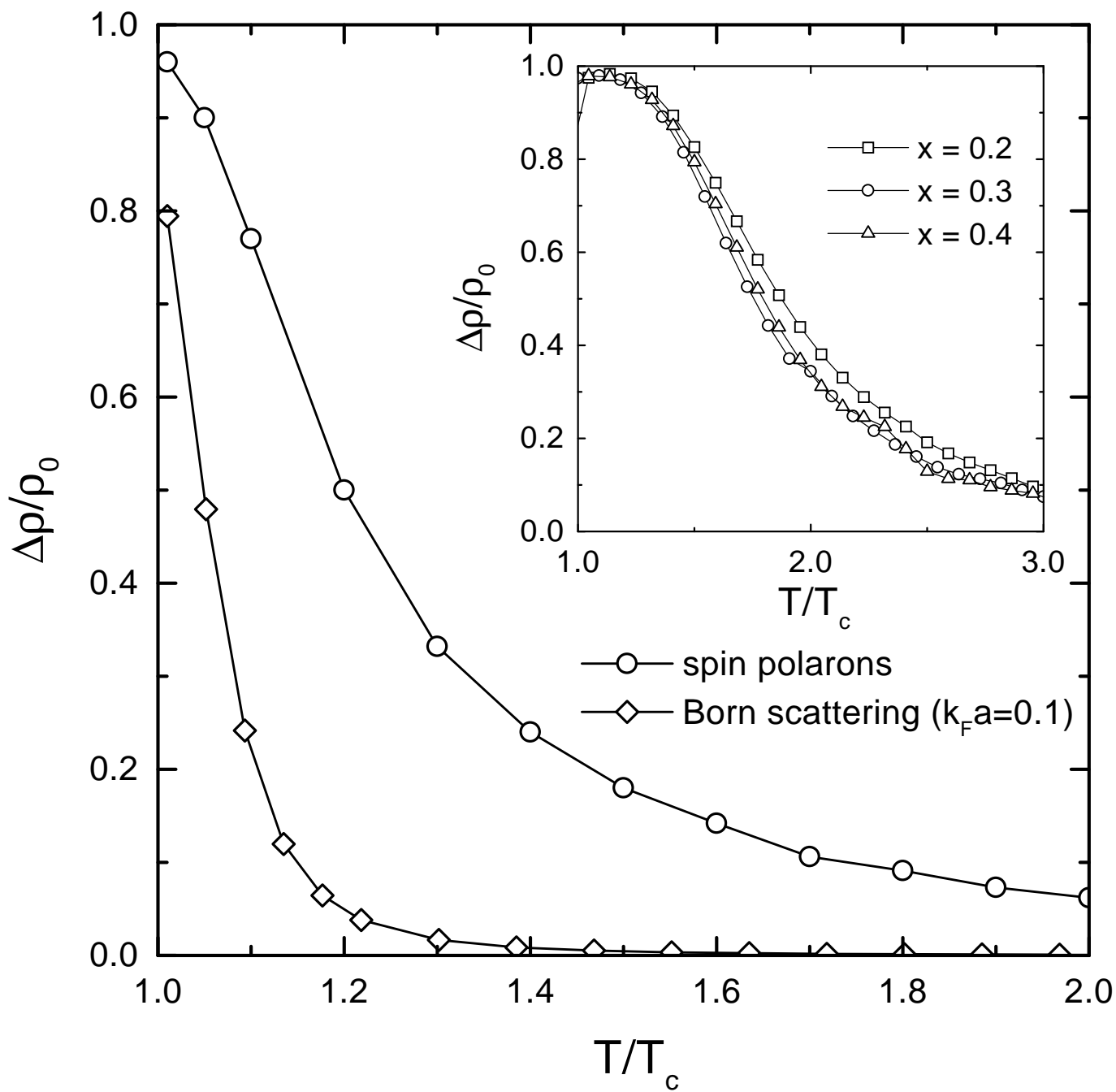
\end{document}